\documentclass[12pt]{article}
\newlength{\mytopmargin}
\newlength{\myleftmargin}
\usepackage{amsmath,amsthm,amssymb,amsbsy,epsfig,graphicx,color,multicol,subfigure}
\usepackage{array,calc}
\usepackage[enableskew]{youngtab}
\usepackage{rotating}
\usepackage{young,epic}
\usepackage{a4wide,bm}
\usepackage{url}
\usepackage{hyperref}

\newtheorem{prop}{Proposition}

\newtheorem{cor}{Corollary}

\usepackage{amsmath,amsfonts,amssymb}

\usepackage{graphicx}

\begin{document}
%
%\begin{frontmatter}

\title{Eigenvalue statistics for product complex Wishart matrices}
\author{Peter J. Forrester${}^\dagger$ }
\date{}
\maketitle
\noindent
\thanks{\small ${}^\dagger$Department of Mathematics and Statistics, 
The University of Melbourne,
Victoria 3010, Australia email:  p.forrester@ms.unimelb.edu.au }

\begin{abstract}
\noindent The eigenvalue statistics for complex $N \times N$ Wishart matrices $X_{r,s}^\dagger X_{r,s}$, where $ X_{r,s}$ is
equal to the product of $r$ complex Gaussian matrices, and the inverse of $s$ complex Gaussian matrices, are considered.
In the case $r=s$ the exact form of the global density is computed. The averaged characteristic polynomial for the corresponding
generalized eigenvalue problem is calculated in terms of a particular generalized hypergeometric function ${}_{s+1} F_r$.
For finite $N$ the eigenvalue probability density function is computed, and is shown to be an example of a biorthogonal ensemble.
A double contour integral form of the corresponding correlation kernel is derived, which allows the hard edge scaled limit
to be computed. The limiting kernel is given in terms of certain Meijer G-functions, and is identical to that found in the
recent work of Kuijlaars and Zhang in the case $s=0$. Properties of the kernel and corresponding correlation functions are
discussed.
\end{abstract}

\section{Introduction}
Let $X$ be a $p \times n$ $(p \ge n)$ matrix with standard complex  Gaussian entries. The positive definite matrix $X^\dagger X$ is then
referred to as a complex Wishart matrix. Such matrices are fundamental in random matrix theory. Their numerous applications
range from the study of the spectrum of random Dirac operators, the eigenvalue statistics of the Wigner-Smith delay time matrix,
and the entanglement of a random pure quantum state in theoretical physics, to the computation of the information capacity
in certain wireless communication systems and the condition number of a linear system of equations in numerical linear algebra,
amongst other examples.
Details and references are given in  \cite[Ch.~3]{Fo10}. 

Crucial to these applications is the exact solvability of statistical properties of the eigenvalues of complex Wishart matrices.
The eigenvalues form a determinantal point process, meaning that the corresponding $k$-point correlation functions
are of the form
\begin{equation}\label{D}
\rho_{(k)}(x_1,\dots,x_k) = \det [ K(x_i,x_j) ]_{i,j=1,\dots,k}
\end{equation}
for $K(x,y)$ independent of $k$. Moreover $K$ itself, referred to as the correlation kernel,
is given by a sum over Laguerre polynomials, making it one of only
four unitary invariant ensembles which can be constructed entirely out of the classical orthogonal polynomials
(see \cite[\S 5.4]{Fo10}). This allows the scaled limit near the spectrum edges, and in the bulk, to be calculated. In the
applied problems listed above, it is typically the scaling limits of the one and two point functions which relate to
observable  quantities. 

Very recently the works \cite{AKW13,AIK13,ABKN13} have extended the class of complex matrices $X$ giving rise to a determinantal
point process for the eigenvalues of $X^\dagger X$. In particular, in \cite{AIK13}  the explicit form of $K(x,y)$ in the case that $X=X_r$, where
\begin{equation}\label{XG}
X_r = G_r G_{r-1} \cdots G_1
\end{equation}
with each $G_k$ a rectangular standard complex Gaussian matrix of dimension $n_k \times n_{k-1}$ where
$n_{0} := N$ and $n_k = N + \nu_k$ $\nu_k \ge 0$, ($k=1,\dots,r$) has been given. This
advancement was soon after complimented by two works: \cite{KZ13} and \cite{Ne13}. In  \cite{KZ13} the hard edge
scaled limit of the correlation kernel was computed (the hard edge scaling refers to the neighbourhood of the spectrum edge at $\lambda = 0$,
scaled so that the spacing between eigenvalues is of order unity).
In \cite{Ne13} knowledge of the averaged characteristic polynomial was used to determine variables in which
the functional form of the global density (scaling such that the support is finite) is very simple.

The product structure of (\ref{XG}) for $X$ in $X^\dagger X$ gives rise to the notion of product  Wishart matrices,
as used in the title of this article. As first noted in \cite{Mu02} in the case $r=2$, and later for general $r$ by the same author
\cite{Mu07}  (see also \cite{AKW13}), product complex Wishart matrices apply in a telecommunications setting when different channels are transmitted
and received via an array of $r-1$ scatters. In this article we extend the class of product  Wishart
matrices from  (\ref{XG}), to also involving the inverse of a product
of standard complex  Gaussian matrices. Thus we set  $X= X_{r,s}$ with
\begin{equation}\label{XG1}
X_{r,s} =      G_rG_{r-1} \cdots G_1     ( \tilde{G}_s \tilde{G}_{s-1} \cdots \tilde{G}_1 )^{-1} .
\end{equation}
Here each $\tilde{G}_k$ is of dimension $\tilde{n}_k \times \tilde{n}_{k-1}$ with
$\tilde{n}_0 = \tilde{n}_s = N$ and $\tilde{n}_k = N + \tilde{\nu}_k$,  $\nu_k \ge 0$, ($k=1,\dots,r$ )
In a scattering setting, the inverse of such a product could possibly arise as the cumulative effect of backscattering. However our
interest in this work is not to explore applied settings, but rather to study the exact solvability properties of the eigenvalue statistics of
this class of product Wishart matrices; of course one hopes that the results presented will be used in applications before too long.

The case $r=s=1$ is already in the literature. Let $C = A^{-1} B$ where $A$ and $B$ are standard complex  Gaussian matrices of
dimensions $M \times M$ and $M \times N$ respectively, with $M \ge N$.  A straightforward calculation detailed in \cite{GN99} in the
real case, extended in \cite[Exercises 3.6 q.~3]{Fo10} to include the complex case, shows that the distribution of $C$ is proportional to
\begin{equation}\label{XG2}
{1 \over \det ( \mathbb I_{N} + C^\dagger C)^{2 N}}.
\end{equation}
One line of study would be to analyze the statistical properties of the non-Hermitian matrix $C$ in the complex plane. This has been done in
a slightly more general setting in \cite{FF11}. Equally (\ref{XG2}) enables the study of the eigenvalue distribution of the product Wishart matrix
$Y = C^\dagger C$. Thus by making the change of variables to $Y$, which introduces a factor $(\det Y)^\alpha$,
$\alpha = M - N$, as the Jacobian \cite[Prop.~3.2.7]{Fo10},
then changing variables to the eigenvalues $\{y_j\}_{j=1,\dots,N}$ and eigenvectors of the Hermitian matrix $Y$ \cite[Prop.~1.3.4]{Fo10}, we see that the
eigenvalue distribution has the explicit functional form proportional to
\begin{equation}\label{XG3}
\prod_{l=1}^N {y_l^\alpha \over (1 + y_l)^{2 N}} \prod_{1 \le j < k \le N} (y_k - y_j)^2, \quad y_l \ge 0 \: \: (l=1,\dots,N).
\end{equation}
This becomes more familiar upon the change of variables $\lambda_l = 1/(1 + y_l)$, as it then reads
\begin{equation}\label{XG4}
\prod_{l=1}^N \lambda_l^\alpha (1 -  \lambda_l)^{\alpha} \prod_{1 \le j < k \le N} (\lambda_k - \lambda_j)^2, \quad 0 \le \lambda_l \le 1 \: \: (l=1,\dots,N),
\end{equation}
which is immediately recognised as an example of a Jacobi unitary ensemble \cite[\S 3.8]{Fo10} and as such the corresponding
statistical properties can be analysed in great detail. We remark that the distribution of $(A^\dagger A)^{-1} (B^\dagger B)$ is called the matrix
$F$-distribution in mathematical statistics \cite{GN99}.

For general $r,s \ge 0$ the non-Hermitian matrix (\ref{XG1}), in the case that all matrices in the product are square, has recently been
considered in \cite{ARRS13}.  The corresponding eigenvalues have been shown to give rise to a rotationally invariant
determinantal point process in the complex plane with an explicit weight function given in terms of the Meijer G-function. This provides a strong
hint that the product Wishart matrix formed out of (\ref{XG1}) enjoys special exact solvability properties, generalizing those already
apparent in the case $r=s=1$ from (\ref{XG4}), and generalizing too those found recently in the case $s=0$, general $r \in \mathbb Z^+$
in  \cite{AKW13,AIK13}.

In Section 2 we take up the problem of determining the global density. Here we are able to determine the explicit functional form in the case
$r=s$. In Section 3 we calculate the characteristic polynomial of the generalized eigenvalue problem associated with the product Wishart matrix
corresponding to (\ref{XG1}). The topic of Section 4 is the computation of the joint eigenvalue PDF for the product Wishart matrix, while
in Section 5 the corresponding correlation functions are calculated and the explicit form of the hard edge scaling is obtained. Asymptotic forms
associated with the scaled kernel and correlation functions are also discussed.

\section{The global density}
 \setcounter{equation}{0}
The theory of free probability (see \cite{Vo00}, \cite{NS06}, \cite{No12} for reviews) provides a powerful calculus for the computation of the global
eigenvalue densiities of sums or products of random matrices, given the global densities of the individual matrices. By the global
density we mean the leading large $N$ form, divided by $N$ to have unit mass, and with the eigenvalues scaled so that the limit is
an order one quantity. For example, for a complex Wishart matrix $X^\dagger X$ with $X$ an $M \times N$ $(M \ge N)$ standard Gaussian matrix, we
must scale the eigenvalues by dividing $X^\dagger X$ by $N$. With $M - N$ fixed the large $N$ leading eigenvalue support is then the
interval $[0,4]$, and the global density is given by the Marchenko-Pastur law
\begin{equation}\label{MP}
\rho_{(1)}^{X^\dagger X}(y) = {1 \over \pi y^{1/2}} \Big ( 1 - {y \over 4} \Big )^{1/2}, \qquad 0 < y < 4.
\end{equation}
In fact it is not the global densities themselves that are the central objects of free probability calculus, but rather certain transforms..

The most fundamental of these is the Stieltjes transform (a type of Green function),
\begin{equation}\label{I.1}
G_Y(z) = \int_I {\rho_{(1)}^Y(y) \over y - z} \, dy, \qquad z \notin I,
\end{equation}
where $I$ denotes the interval of support. For (\ref{MP}) we have
\begin{equation}\label{I.2}
G_{X^\dagger X}(z) = {-1 + \sqrt{1 - 4/z} \over 2}
\end{equation}
(see e.g.~\cite[Exercises 14.4 q.6(i) with $\alpha = 0$]{Fo10}). And since the eigenvalues of $(X^\dagger X)^{-1}$ are the reciprocals of the
eigenvalues of $X^\dagger X$, a straightforward calculation from (\ref{I.1}) and (\ref{I.2}) shows
\begin{equation}\label{I.3}
G_{(X^\dagger X)^{-1}}(z) = - {1 \over z} - {-1 + \sqrt{1 - 4z} \over 2z^2},
\end{equation}
this being a special case of the general relation
\begin{equation}\label{GI}
G_{Y^{-1}}(z) = - {1 \over z} - {G_Y(1/z) \over z^2}.
\end{equation}

Now introduce the auxiliary quantity
\begin{equation}\label{EG}
\Upsilon_Y(z) := - 1 - {G(1/z) \over z}
\end{equation}
so that
$$
\Upsilon_{X^\dagger X}(z)  = - 1 - \Big ( {-1 + \sqrt{1 - 4 z} \over 2 z} \Big ), \quad
\Upsilon_{(X^\dagger X)^{-1}}(z)  = z \Big ( {-1 + \sqrt{1 - 4 / z} \over 2 } \Big ).
$$
From these explicit forms we compute the corresponding inverse functions
\begin{equation}\label{E1}
\Upsilon_{X^\dagger X}^{-1}(z) = {z \over (1 + z)^2}, \qquad \Upsilon_{(X^\dagger X)^{-1}}^{-1}(z) =  - {z^2 \over 1 + z}.
\end{equation}
Finally, introduce the $S$-transform by
\begin{equation}\label{4E}
S_Y(z) = {1 + z \over z}  \Upsilon_Y^{-1}(z).
\end{equation}
We see from (\ref{E1}) that
\begin{equation}\label{4.0}
S_{X^\dagger X}(z) = {1 \over 1 + z}, \qquad S_{(X^\dagger X)^{-1}}(z) = - z.
\end{equation}

The key feature of the $S$-transform is that for $A$ and $B$ free random matrices --- the case that $A$ and $B$
are independent Gaussian random matrices has this property \cite[pg.~90]{Mu07} --- one has  \cite[pg.~99]{Mu07}
\begin{equation}\label{4.1}
S_{AB}(z) = S_A(z) S_B(z).
\end{equation}
By using (\ref{4.1}) and (\ref{4.0}) we can compute the $S$ transform of the product Wishart matrix corresponding to (\ref{XG1}).

\begin{prop}
Let $X_{r,s}$ be given by (\ref{XG1}). Suppose the matrices $\tilde{G}_1,\dots, \tilde{G}_s$ have size $N \times N$, and that to leading
order the size of the matrices $G_1,\dots,G_r$ is $N \times N$. Suppose furthermore that each of the matrices is divided by $N$ so that
the global densities have Stieltjes transforms (\ref{I.1}) or (\ref{I.2}) as appropriate. We have that the 
Stieltjes transform of $X_{r,s}^\dagger X_{r,s}$ satisfies the functional equation
\begin{equation}\label{GG}
\Big ( 1 - {G(-1/z) \over z} \Big )^{s+1} = z  \Big (  {G(-1/z) \over z} \Big )^{r+1}.
\end{equation}
\end{prop}

\noindent Proof. \quad Our strategy is based on the working in \cite{Mu07} in which the case $s=0$ is
considered. The eigenvalues of the product Wishart matrix $X_{r,s}^\dagger X_{r,s}$ are identical to the eigenvalues of
$(X_{r,s-1}^\dagger X_{r,s-1}) (\tilde{G}_s^\dagger \tilde{G}_s)^{-1}$. Applying (\ref{4.1}) and the second equation in (\ref{4.0})
to the latter it follows that
$$
S_{X_{r,s}^\dagger X_{r,s}}(z) = (-z)  S_{X_{r,s-1}^\dagger X_{r,s-1}}(z).
$$
Now iterating this shows
$$
S_{X_{r,s}^\dagger X_{r,s}}(z) = (-z)^s  S_{X_{r,0}^\dagger X_{r,0}}(z).
$$

For notational convenience, let us now relabel $G_r G_{r-1} \cdots G_1$  to read $G_1 G_2 \cdots G_r$. With this done, 
we note that $X_{r,0}^\dagger X_{r,0}$ has the same eigenvalues as $(X_{r-1,0}^\dagger X_{r-1,0})(G_r G_r^\dagger)$. 
Noting that $G_r G_r^\dagger$ have the same nonzero eigenvalues as $G_r^\dagger G_r$ and applying the first equation
in (\ref{4.0}), (\ref{4.1}), and iterating we conclude
$$
S_{X_{r,s}^\dagger X_{r,s}}(z)  = { (-z)^s \over (1 + z)^r}.
$$
Recalling (\ref{4E}) it follows from this that
$$
z = (-1)^s {( \Upsilon_{X_{r,s}^\dagger X_{r,s}}(z) )^{s+1} \over (1 +  \Upsilon_{X_{r,s}^\dagger X_{r,s}}(z) )^{r+1} }.
$$
Now recalling (\ref{EG}) and performing minor manipulation, (\ref{GG}) follows. \hfill $\square$

\medskip
We remark that (\ref{GG}) is unchanged by the mappings
$$
r \leftrightarrow s, \quad
G(z) \mapsto - {1 \over z} - {G(1/z) \over z^2}, \quad z \mapsto {1 \over z},
$$
which is in keeping with the general relation (\ref{GI}) and the structure of $X_{r,s}^\dagger X_{r,s}$. Another general property
of (\ref{GG}) is that it implies the first moment $m_1^{X_{r,s}^\dagger X_{r,s}}$ must be infinite and thus that the global density
must be supported on a semi-infinite interval for $s  \ge 1$. To see this, suppose to the contrary that the first moment is finite. Then we can
expand $G(-1/z)/z = 1 - z m_1^{X_{r,s}^\dagger X_{r,s}} + O(z^2)$. Substituting in (\ref{GG}) gives that to leading order in $z$,
$z^{s+1} (m_1^{X_{r,s}^\dagger X_{r,s}} )^{s+1})= z$, which is only consistent for $s=0$.

The case $s=0$ of (\ref{GG}) is well known \cite{Mu07,AGT10,BBCC11,BJLNS11}. Expanding the Green function in terms of the moments $m_p^{X_r^\dagger X_r}$
of the density gives the recurrence
\begin{equation}
m_p^{X_r^\dagger X_r} = \sum_{q_1,\dots,q_{r+1} \ge 0 \atop q_1 + \cdots q_{r+1} = p -1 } m_{q_1} \cdots m_{q_{r+1}}.
\end{equation}
Together with the initial condition $m_0^{X_r^\dagger X_r} =1$, the unique solution of this recurrence is the Fuss-Catalan
numbers, giving that
\begin{equation}\label{U}
m_p^{X_r^\dagger X_r} = \bigg ( {r p + p \atop p} \bigg ) {1 \over r p + 1}.
\end{equation}
Very recently \cite{Ne13}, upon the introduction of the variable $\phi$ according to
\begin{equation}\label{B1}
x = {(\sin (r + 1) \phi )^{r+1} \over \sin \phi (\sin r \phi)^r }, \qquad 0 < \phi < {\pi \over r+1}
\end{equation}
it has been shown that the corresponding eigenvalue density is given by the succinct expression
\begin{equation}\label{B2}
\rho_{(1)}^{X_r^\dagger X_r} (\phi) = {(\sin \phi)^2 (\sin r \phi)^{r-1} \over \pi (\sin (r+1) \phi)^r}.
\end{equation}
A multiple integral formula has been given in \cite{LSW11}, and there is also a formula in terms
of a sum of $r$ generalized hypergeometric functions \cite{PZ11}. Of particular interest is the
singular behaviour in the original variable $x$ as $x \to 0^+$,
\begin{equation}\label{B3}
\rho_{(1)}^{X_r^\dagger X_r} (x) \sim { \sin \pi/(r+1) \over \pi x^{r/(r+1)}},
\end{equation}
which follows from (\ref{B1}) and (\ref{B2}).

Another case of (\ref{GG}) which allows for explicit determination of the density is when $r=s$.

\begin{cor}
In the case $r=s$ the global density is supported on $(0,\infty)$ and has the explicit form
\begin{equation}\label{C1}
x \rho_{(1)}^{X_{r,r}^\dagger X_{r,r}} (x) = {1 \over \pi} {x^{1/(r+1)} \sin \pi/(r+1) \over
1 + 2 x^{1/(r+1)} \cos \pi/(r+1) +  x^{2/(r+1)} }.
\end{equation}
\end{cor}

\noindent Proof. \quad We have from (\ref{GG}) in the case $r=s$ that
$$
{z G_{X^\dagger_{r,r} X_{r,r}}(-z) \over 1 - z  G_{X^\dagger_{r,r} X_{r,r}}(-z) } = z^{1/(r+1)},
$$
and thus
$$
z G_{X^\dagger_{r,r} X_{r,r}}(-z)  = 1 - {1 \over 1 + z^{1/(r+1)}}.
$$
From the definition (\ref{I.1}) it follows from this that
\begin{equation}\label{Ma}
\int_I {\lambda  \rho_{(1)}^{X_r^\dagger X_r} (\lambda) \over \lambda + z} \, d \lambda =
{1 \over 1 + z^{1/(r+1)}}.
\end{equation}
Applying the inverse formula
$$
x  \rho_{(1)}^{X_r^\dagger X_r} (x)  = - {1 \over 2 \pi i} \bigg (
{1 \over 1 + z^{1/(r+1)} } \Big |_{z = x e^{\pi i}} -  {1 \over 1 + z^{1/(r+1)} } \Big |_{z = x e^{-\pi i}} \bigg )
$$
gives (\ref{C1}). \hfill $\square$

\medskip

Changing variables $\lambda = 1/(1 + x)$ transforms the density to have support on $(0,1)$.
It follows from (\ref{Ma}) that the transformed density satisfies
$$
{1 \over z} \Big ( 1 - {1 \over 1 + z^{1/(r+1)}}\Big ) =  \int_0^1 { \lambda  \rho_{(1)}^{X_{r,r}^\dagger X_{r,r}} (\lambda) \over
1 - (1 - z) \lambda}\, d \lambda,
$$
and thus that $p$-th moment is equal to the coefficient of $(1-z)^{p-1}$ in the power series expansion about $z=1$ of the LHS. However,
unlike the moments (\ref{U}), these moments don't appear to have any further structure except in the case $r=s=1$
when, as can be checked from (\ref{C1}), the transformed density is equal to the particular beta density
$$
 \rho_{(1)}^{X_1^\dagger X_1} (\lambda) = {1 \over \pi} {1 \over \sqrt{\lambda (1 - \lambda)}}, \qquad 0 < \lambda < 1.
 $$
 
 We remark that the $x \to 0^+$ leading form of (\ref{C1}) is exactly the same as that exhibited by (\ref{B3}) in the
 case $s=0$, suggesting this to be  a universal feature valid for general $r,s$ which is independent of $s$.
 
 \section{The characteristic polynomial}\label{S3}
 \setcounter{equation}{0}
 The averaged characteristic polynomial $P_N^{X_r^\dagger X_r}(x)$ for the product complex Wishart ensemble in the case $s=0$ has
 been computed in \cite{AIK13}. Introduce the  generalized hypergeometric function
 \begin{equation}\label{pFq}
 {}_p F_q  \Big ( { a_1, \dots, a_p  \atop  b_1, \dots,  b_q} \Big | x \Big ) = \sum_{k=0}^\infty {(a_1)_k \cdots (a_p)_k \over
(b_1)_k \cdots (b_q)_k } {x^k \over k!},
\end{equation}
 where $(c)_k := c(c+1) \cdots (c+k-1)$. Then with
 $\nu_j := n_{j} - N$ we have
 \begin{equation}\label{KP}
 P_N^{X_r^\dagger X_r}(x) = (-1)^N \prod_{l=1}^r (\nu_l + 1)_N \:
 {}_1 F_r \Big ( {-N \atop \nu_1+1, \dots, \nu_r+1} \Big | x \Big ).
 \end{equation}
 
 For product Wishart matrices formed from (\ref{XG1}) the averaged characteristic polynomial is not well defined due to the
 divergence caused by the inverse matrices. To avoid this, we consider the associated generalized eigenvalue problem.
 Explicitly, with
 \begin{equation}\label{ATG}
 A_r := G_r G_{r-1} \cdots G_1, \qquad \tilde{A}_s = \tilde{G}_s \tilde{G}_{s-1} \cdots \tilde{G}_1
 \end{equation}
 so that $X_{r,s} = A_r \tilde{A}_s^{-1}$, we consider the polynomial
 \begin{equation}\label{PA}
 P_N^{(r,s)}(\lambda) := \Big \langle \det( \lambda \tilde{A}_s^\dagger \tilde{A}_s - A_r^\dagger A_r ) \Big  \rangle,
 \end{equation}
where the average is over the random Gaussian matrices  $\{A_j\}$, $\{\tilde{A}_k \}$. Note that (\ref{PA}) reduces to the
characteristic polynomial $ P_N^{X_r^\dagger X_r}(x)$ in the case $s=0$.

A fundamental insight from \cite{AIK13} (see also \cite{IK13})
 is that the eigenvalue PDF of the product Wishart matrix $X_r^\dagger X_r$ with $X_r$ given by
 (\ref{XG}) is the same as that in which each $G_j$ is an $N \times N$ complex Gaussian matrix with distribution
 proportional to
 \begin{equation}\label{E1a} 
 (\det G_j^\dagger G_j)^{\nu_j} e^{- {\rm Tr} \,  G_j^\dagger G_j}
 \end{equation} 
 (recall that in (\ref{XG}) the size of $G_j$ is $n_j \times n_{j-1}$, with $n_j - N = \nu_j \ge 0$ and $n_0 = N$).
  We know from \cite{FBKSZ12} that this distribution can be realized by forming the random matrix
 $ ({H}_j^\dagger {H}_j)^{1/2} U$, where $U$ is a random unitary matrix and ${H}_j$ is 
 an $n_j \times N$ standard complex Gaussian matrix.
 Similarly, we can replace the rectangular matrices $\tilde{G}_l$ in (\ref{ATG}) by $N \times N$ matrices with
 distribution proportional to
 \begin{equation}\label{E1am} 
 (\det \tilde{G}_l^\dagger \tilde{G}_l)^{\mu_l} e^{- {\rm Tr} \,  \tilde{G}_l^\dagger \tilde{G}_l},
 \end{equation} 
 or equivalently replace each $\tilde{G}_l$ by  $ (\tilde{H}_l^\dagger \tilde{H}_l)^{1/2} U$ where 
 $\tilde{H}_l$ is an $\tilde{n}_l \times N$ standard complex Gaussian matrix.
 Although not used in our study, we note from \cite{IK13} that changing the order of the products of these $N \times N$ matrices
 in the definition of $X_{r,s}$ does not change the corresponding eigenvalue PDF.

With the product now consisting exclusively of $N \times N$ matrices, as with (\ref{KP}) we can evaluate the polynomial (\ref{PA}) in terms of the generalized
hypergeometric function (\ref{pFq}).

\begin{prop}\label{P2}
We have
\begin{equation}\label{Prs}
 P_N^{(r,s)}(\lambda) =(-1)^N \prod_{l=1}^r (\nu_l + 1)_N
 \, {}_{s+1} F_r \Big ( {-N,-\mu_1- N, \dots, - \mu_s -N \atop
 \nu_1+1, \dots, \nu_r+1} \Big | (-1)^s \lambda \Big ).
 \end{equation}
 \end{prop}
 
 \noindent
 Proof. \quad We require the notion of a Schur polynomial, and some associated formulas.
 With $\kappa$ a partition defined by its parts $\kappa_1 \ge \kappa_2 \ge \cdots \ge \kappa_N$, and
 $\{y_j\}$ denoting the eigenvalues of the matrix $Y$, we define the
 Schur polynomial $s_\kappa(y_1,\dots,y_N) =: s_\kappa(Y)$ by the ratio of determinants
 $$
 s_\kappa(y_1,\dots,y_N) = {\det [ y_j^{\kappa_k + N - k} ]_{j,k=1,\dots,N} \over
 \det [ y_j^{N-k}]_{j,k=1,\dots,N}}.
 $$
 Since, with $(1^k)$ the partition consisting of $k$ 1's and all other parts zero,
 \begin{equation}\label{se}
 s_{( 1^k)}(y_1,\dots, y_N) = e_k (y_1,\dots,y_N),
  \end{equation}
  where $e_k$ denotes the $k$-th elementary symmetric function (polynomial), we have
 \begin{equation}\label{F2}
 \det ( \lambda \mathbb I_N - Y) = (-1)^N \sum_{k=0}^N (-\lambda)^k  s_{(1^{N-k})}(Y).
  \end{equation}
  
  A key formula relating to the Schur polynomial is that it factorizes upon averaging over the unitary group
  (see e.g.~\cite{FR09} and references therein)
 \begin{equation}\label{SAB}
 \langle s_\kappa(A U^\dagger B U) \rangle_U = {s_\kappa(A) s_\kappa(B) \over s_\kappa(\mathbb I_N)}.
 \end{equation}
We require too the fact that
 \begin{equation}\label{SAB1}      
 (\det Y)^p s_\kappa(Y) = s_{\kappa + p^N}(Y),
 \end{equation}
 where $\kappa + p^N := (\kappa_1+p,\dots, \kappa_N + p)$. This allows us to give meaning to a Schur polynomial
 indexed by partitions with negative parts, and in particular to the identity \cite[Exercises 12.1 q.2]{Fo10}
 \begin{equation}\label{SAB2}   
 s_\kappa(Y^{-1}) = s_{- \kappa^R}(Y),
 \end{equation}
 where $\kappa^R := (\kappa_N,\kappa_{N-1},\dots,\kappa_1)$. 
 
 To make use of the above theory, we rewrite (\ref{PA}) to read
 \begin{equation}\label{F1}
   P_N^{(r,s)}(\lambda) = 
    \Big \langle \det  \tilde{A}_s^\dagger \tilde{A}_s \,
 \det( \lambda \mathbb I_N - ( \tilde{A}_s^\dagger \tilde{A}_s)^{-1} A_r^\dagger A_r ) \Big  \rangle .   
 \end{equation}
 We can now expand the second determinant according to (\ref{F2}). Once having done this, we note that each average
 is unchanged by the replacement of $A_r$ by $A_r U$ with $U$ a unitary matrix. This means that we can average over $U$
 without changing the values of the original averages. The average over $U$ gives a factorization of the Schur polynomials
 according to (\ref{SAB}), and we conclude
  \begin{align}\label{ST}
     P_N^{(r,s)}(\lambda) &=  (-1)^N \sum_{k=0}^N { (-\lambda)^k \over s_{(1^{N-k})}(\mathbb I_N)}
    \Big \langle \det  \tilde{A}_s^\dagger \tilde{A}_s \, s_{(1^{N-k})}((\tilde{A}_s^\dagger \tilde{A}_s)^{-1}) \Big \rangle
   \Big \langle     s_{(1^{N-k})} (A_r^\dagger A_r)  \Big \rangle \nonumber \\
   & = (-1)^N \sum_{k=0}^N { (-\lambda)^k \over s_{(1^{N-k})}(\mathbb I_N)}
    \Big \langle s_{(1^{k})}(\tilde{A}_s^\dagger \tilde{A}_s) \Big \rangle
   \Big \langle     s_{(1^{N-k})} (A_r^\dagger A_r)  \Big \rangle ,
\end{align}   
where the second equality follows upon use of (\ref{SAB1}) and (\ref{SAB2}). Recalling the definition (\ref{ATG}) of 
$A_r$ and $\tilde{A}_s$, we see that by the introduction of unitary matrices and use of (\ref{SAB}) as described above
all the averages in (\ref{ST}) can be factorized into averages over single Wishart matrices,
 \begin{equation}\label{SAV}
  P_N^{(r,s)}(\lambda)  = \sum_{k=0}^N   (-\lambda)^k s_{(1^{k})}(\mathbb I_N) 
  \prod_{j=1}^s { \langle s_{(1^{k})}(\tilde{G}_j^\dagger \tilde{G}_j)\rangle \over  s_{(1^{k})}(\mathbb I_N) }
  \prod_{l=1}^r  { \langle s_{(1)^{N-k}}(G_l^\dagger G_l) \rangle \over  s_{(1)^{N-k}}(\mathbb I_N) }.
  \end{equation}
  
  The averages in (\ref{SAV}) are well known. 
  After recalling the distributions (\ref{E1a}) and (\ref{E1am}),
  changing variables to the Wishart matrices $\tilde{X}_j := \tilde{G}_j^\dagger \tilde{G}_j $,
  then changing variables to the
  eigenvalues and eigenvectors we can make use of a (a special case of) the Kadell-Kaneko-Macdonald integration formula
  \cite[eq.~(12.152)]{Fo10} to conclude
  $$
   { \langle s_{(1^{k})}(\tilde{G}_j^\dagger \tilde{G}_j \rangle \over  s_{(1^{k})}(\mathbb I_N) } = (-1)^k (-\mu_j -N)_k, \qquad
 { \langle s_{(1)^{N-k}}(G_l^\dagger G_l) \rangle \over  s_{(1)^{N-k}}(\mathbb I_N) } = {(\nu_l + N)! \over \nu_l !} {1 \over (\nu_l + 1)_k}.
 $$
 Furthermore, it follows from (\ref{se}) that $ s_{(1^{k})}(\mathbb I_N)  = \binom{N}{k}$. Substituting these formulas
 in (\ref{SAV}) and comparing with the definition (\ref{pFq}) gives (\ref{Prs}). \hfill $\square$
 
 We note that the formula (\ref{Prs}) reduces to (\ref{KP}) in the case $s=0$, as it must. Furthermore, we can check the functional
 property
 $$
 (-1)^N \lambda^N P^{(r,s)}_N( {1 \over \lambda} ) = P_N^{(s,r)}(\lambda) \Big |_{\{\mu_j\} \leftrightarrow \{\nu_k\}}
 $$
 which is evident from the definition (\ref{PA}).

 \section{Eigenvalue PDF for product complex Wishart matrices}
 \setcounter{equation}{0}
 In keeping with the discussion above Proposition  \ref{P2}, our problem is equivalent to
 considering products of independent $N \times N$ matrices, with the joint distribution of these
 matrices proportional to 
 \begin{equation}\label{E3}  
 \prod_{j=1}^r (\det  G_j^\dagger G_j)^{\nu_j}   e^{- {\rm Tr} \,  G_j^\dagger G_j} 
 \prod_{l=1}^s   (\det \tilde{G}_l^\dagger \tilde{G}_l)^{\mu_l} e^{- {\rm Tr} \,  \tilde{G}_l^\dagger \tilde{G}_l}.
 \end{equation} 
 Given this, and with $X^{(r,s)}$ specified in terms of $\{G_j\}$, $\{\tilde{G}_l\}$ as in (\ref{XG1}), we seek the
 eigenvalue PDF of the product Wishart matrix $(X^{(r,s)})^\dagger X^{(r,s)}$.
 For this we require the Meijer $G$-function
 \begin{equation}\label{GGa}
G_{p,q}^{m,n} \Big ( z \Big | {a_1,\dots, a_p \atop b_1,\dots, b_q} \Big )
=
{1 \over 2 \pi i}
\int_C {\prod_{j=1}^m \Gamma ( b_j - s) \prod_{j=1}^n \Gamma(1 - a_j + s) \over
\prod_{j=m+1}^q \Gamma (1-  b_j + s) \prod_{j=n+1}^p \Gamma(a_j - s) } z^s \, ds,
\end{equation}
where $C$ is an appropriate contour as occurs in the inversion formula for the corresponding  Mellin transform
(for a recent review on the Meijer $G$-function see \cite{BS13}).

\begin{prop}\label{P3}
Let the eigenvalues of the above specified product Wishart matrix be denoted $\{x_j\}_{j=1,\dots,N}$,
and suppose $s \ge 1$.
The PDF for these eigenvalues is proportional to
\begin{equation}\label{Box}
\prod_{1 \le j < k \le N} (x_k - x_j) \: \det \Big [ G^{r,s}_{s,r} \Big ({ -(\mu_1 + j - 1 + N),
-(\mu_2 + N), \dots, - (\mu_s+N)  \atop \nu_1,\nu_2,\dots,\nu_r} \Big | x_k \Big ) \Big ]_{j,k=1,\dots,N}.
\end{equation}
\end{prop}

\noindent
Proof. \quad
 We  proceed in an analogous fashion to \cite{AKW13} for the derivation of (\ref{Box}) in the case $s=0$,
 and introduce the square matrices $\{Y_l\}_{l=1,\dots,s}$,
 $ \{Z_j\}_{j=1,\dots,r}$ via the change of variables
 $$
 Y_l =  \tilde{G}_l Y_{l-1} \: \; (Y_0 = \mathbb I_N; \, l=1,\dots, s) \qquad
 Z_j = {G}_{j}  Z_{j-1}  \: \;  (Z_0 = (Y_s)^{-1} \: \: j=1,\dots, r).
 $$ 
 Upon noting that for square complex matrices $A,B,C$ such that $A = B^{-1}C$ with $B$ fixed, we have $(dA) = (\det B)^{-N} (dC)$,
 this gives
 \begin{equation}\label{I1}
 \prod_{l=1}^s (\det Y_l^\dagger Y_l)^{\mu_l - \mu_{l+1} - N} e^{- {\rm Tr} \, Y_l^\dagger Y_l (Y_{l-1}^\dagger Y_{l-1})^{-1}} \,
 \prod_{j=1}^r (\det Z_j^\dagger Z_j)^{\nu_j - \nu_{j+1} - N} e^{- {\rm Tr} \, Z_j^\dagger Z_j (Z_{j-1}^\dagger Z_{j-1})^{-1}} 
 \end{equation}
 where $ \mu_{s+1} =-( 2N + \nu_1)$, $\nu_{r+1} = -N$.

Next we change variables $Y_l^\dagger Y_l = A_l$, $Z_j^\dagger Z_j = B_j$. We see from the definitions that
$B_r = X_{r,s}^\dagger X_{r,s}$. Since the $Y_l$ and $Z_j$ are complex square matrices,
there is no new Jacobian factor in this step, and (\ref{I1}) reads
 \begin{equation}\label{I2}
 \prod_{l=1}^s (\det A_l)^{\mu_l - \mu_{l+1} - N} e^{- {\rm Tr} \, A_l  A_{l-1}^{-1}} \,
 \prod_{j=1}^r (\det  B_j)^{\nu_j - \nu_{j+1} - N} e^{- {\rm Tr} \,  B_j  B_{j-1}^{-1}} 
 \end{equation}
 ($A_0 := \mathbb I$, $B_0 := A_s^{-1}$).
 Changing variables to the eigenvalues and eigenvectors of $\{A_l\}$ and $\{B_j\}$ introduces Jacobian factors
 $\prod_{l=1}^s \Delta(a^{(l)})  \prod_{j=1}^r  \Delta(b^{(j)}) $, where $a^{(l)}$, $b^{(j)}$ denotes the set of
 eigenvalues of $\{A_l\}$ and $\{B_j\}$ respectively, and with $X = \{x_1,\dots x_N\}$, $\Delta(X) := \prod_{1 \le j < k \le N}(x_k - x_j)$.
 Moreover, the integration over the eigenvectors can done by making use of the Harish-Chandra/ Itzykson--Zuber formula
 \cite[Prop.~11.6.1]{Fo10}
 \begin{equation}\label{HC}
 \int e^{{\rm Tr} \, (U A U^\dagger B)} (U^\dagger dU) \propto
 {\det [ e^{a_j b_k} ]_{j,k=1,\dots,N} \over \Delta(\{a_j\}) \Delta(\{b_k\}) }.
 \end{equation}
 Doing this, starting with $B_r$ and simplifying by using the fact that $1/\Delta(\{x^{-1}\}) = \prod_{l=1}^Nx_l^{N-1}/\Delta(\{x\})$ (here we have
 ignored signs) we obtain
 \begin{align}\label{4.7}
 &\Delta (b^{(p)})  \prod_{j=0}^{r-1} \Big (  \prod_{k=1}^N (b_k^{(r-j)})^{\nu_{r-j} - \nu_{r-j+1} - 1}  \Big ) \det [ e^{-b_m^{(r-j)} (b_n^{(r-j-1)})^{-1}}]_{m,n=1,\dots,N} \nonumber \\
 & \times   \prod_{l=0}^{s-1} \Big (  \prod_{k=1}^N (a_k^{(s-l)})^{\mu_{s-l} - \mu_{s-l+1} - 1} \Big )  \det [ e^{-a_m^{(s-l)} (a_n^{(s-l-1)})^{-1}}]_{m,n=1,\dots,N } \,
 \Delta(a^{(1)}),
 \end{align}
 where in this formula we require $\nu_{r+1}=-1$, $\mu_{s+1} = - \nu_1 - N - 1$, $b^{(0)} = (a^{(s)})^{-1}$, $a^{(0)}=1$.

 We now integrate over $b^{(r-1)}, b^{(r-2)}, \dots b^{(1)}, a^{(s)}, a^{(s-1)}, \dots, a^{(1)}$ in order. To do this we make repeated
 use the integration  formula \cite[Eq.~(5.170)]{Fo10}
 \begin{equation}\label{170}
  \int_I dx_1 \cdots \int_I dx_N \, \det [\xi_j(x_k)]_{j,k=1,\dots,N} \det [\eta_j(x_k)]_{j,k=1,\dots,N} 
  = N! \det \Big [ \int_I \xi_j(x) \eta_k(x) \, dx \Big ]_{j,k=1,\dots,N}.
 \end{equation}
 We then obtain, up to constant factors,
  \begin{equation}\label{I3}
 \Delta (b^{(r)}) 
 \det[ I_j ( b^{(r)}_k)]_{j,k=1,\dots,N}
  \end{equation}
 where
  \begin{multline}\label{HC1}
   I_j (b^{(r)}) :=  \int_{[0,\infty]^{r+s-1} }db^{(r-1)} \cdots db^{(1)} da^{(s)} \cdots da^{(1)} \, (a^{(1})^{j-1}
   \prod_{l=1}^s (a^{(l)})^{\mu_l - \mu_{l+1} - 1} e^{- a^{(l)} (a^{(l-1)})^{-1}} \nonumber \\
\times \prod_{j=1}^r (b^{(j)})^{\nu_j - \nu_{j+1} - 1} e^{- b^{(j)} (b^{(j-1)})^{-1}} .
 \end{multline}  
 To simplify the integral we change variables $a^{(l)}/a^{(l-1)} = y_l $, $(l=s,\dots,1)$ with $a^{(l-1)}$ regarded as fixed,
 and $b^{(j)}/ b^{(j-1)} = t_j$, $(j=r-1,\dots,1)$ with $b^{(j-1)}$ regarded as fixed. This shows
 \begin{equation}\label{HC2}
   I_j (b^{(r)}) =  (b^{(r)})^{\nu_r}  \int_{[0,\infty]^{r+s-1}} dt_{r-1} \cdots dt_1 dy_s \cdots dy_1 \, y_1^{j-1}  \prod_{l=1}^s y_l^{\mu_l+N+\nu_r} e^{-y_l}
  \prod_{j=1}^{r-1} t_j^{\nu_j-1-\nu_r} e^{-t_j}   \, e^{-b^{(r)}/b^{(r-1)}}
  \end{equation}
  
  The next step is to write
  \begin{align*}
  e^{-b^{(r)}/b^{(r-1)}} & = \int_0^\infty dt_r \, \delta \Big ( t_r - b^{(r)} {\prod_{l=1}^{s} y_l \over \prod_{j=1}^{r-1} t_j } \Big ) e^{-t_r} \\
  & = {\prod_{l=1}^{r-1} t_l \over \prod_{l=1}^{s} y_l} \,  \int_0^\infty dt_r \,   \delta \Big (  b^{(r)} -  {\prod_{j=1}^{r} t_j \over \prod_{l=1}^{s} y_l}  \Big )
   e^{-t_r} .
  \end{align*}
  The first line in this equation is equivalent to the equation $a^{(0)}/(b^{(0)} a^{(s)}$, which is consistent with the
  requirements noted below  (\ref{4.7}).
  Substituting this in (\ref{HC2}) shows
  \begin{multline}\label{HC3}
   I_j (b^{(r)}) :=     \int_{[0,\infty]^{r+s} }dt_{r} \cdots dt_1 dy_s \cdots dy_1 \, y_1^{j-1}  \prod_{l=1}^s y_l^{\mu_l+N - 1} e^{-y_l}
     \prod_{j=1}^{r} t_j^{\nu_j} e^{-t_j}  \\
     \times  \delta \Big (  b^{(r)} -  {\prod_{j=1}^{r} t_j \over \prod_{l=1}^{s} y_l}  \Big ). 
\end{multline}     
We note that when $j=1$, $N=1$ (\ref{HC3}) is precisely the distribution of the ratio $\prod_{j=1}^{r} t_j /  \prod_{l=1}^{s} y_l$ with
the random variables $\{y_l\}$,  $\{t_j\}$ having PDF proportional to $ y_l^{\mu_l} e^{-y_l}$,  $ t_j^{\nu_j} e^{-t_j}$ respectively.
According to (\ref{I3}), since $N=1$ this case of $I_j(b^{(r)})$ gives the distribution of the random variable $b^{(r)}$, which is
indeed equal to the stated ratio.

Our remaining task is to evaluate (\ref{HC3}) in terms of the Meijer $G$-function.
For this we see from (\ref{HC3}) that the Mellin transform of $I_j$ is very simple. Thus we have
$$
\int_0^\infty b^{s-1} I_j(b) \, db =
(\prod_{l=1}^r \Gamma (\nu_l + s)) \Gamma(\mu_1 + N - s + j)  \prod_{l=2}^{s} \Gamma(\mu_l + N - s + 1).
$$
Taking the inverse Mellin transform implies
\begin{equation}\label{MT}
I_j(b) = {1 \over 2 \pi i} \int_{c- i \infty}^{c + i \infty} \Big ( \prod_{l=1}^r \Gamma (\nu_l - s) \Big ) \Gamma(\mu_1 + N +s + j)  \prod_{l=2}^{s} \Gamma(\mu_l + N + s + 1)
b^s \, ds,
\end{equation}
where $c>0$.
Comparison with the definition (\ref{GGa}) shows that this is a particular Meijer $G$-function. Substituting in (\ref{I3}) gives the
stated formula (\ref{Box}). \hfill $\square$

\medskip
Instead of requiring that $s \ge 1$, we can reformulate (\ref{Box}) so that we require $r \ge 1$.

\begin{cor}
In the setting of Proposition \ref{P3} suppose $r \ge 1$. Up to a proportionality constant, the eigenvalue PDF can be written
\begin{equation}\label{Box1}
\prod_{1 \le j < k \le N} (x_k - x_j) \: \det \Big [ G^{r,s}_{s,r} \Big ({ -(\mu_1 + N), \dots
, - (\mu_s+N)  \atop \nu_1+j-1,\nu_2,\dots,\nu_r} \Big | x_k \Big ) \Big ]_{j,k=1,\dots,N}.
\end{equation}
\end{cor}

\noindent
Proof. \quad From the set up of the problem, the eigenvalue PDF must be unchanged by the change of variables
$x_j \mapsto 1/x_j$ provided we make the interchange $\{\mu_l\}_{l=1,\dots,s} \leftrightarrow \{\nu_j\}_{j=1,\dots,r}$.
Doing this, with the aid of standard functional properties of the Meijer $G$-function, (\ref{Box1}) follows from (\ref{Box}).
\hfill $\square$

\medskip
The case $s=0$ of (\ref{Box1}) was derived in \cite{AIK13}.

\section{Correlation functions}
 \setcounter{equation}{0}
 \subsection{Finite $N$}
 The eigenvalue PDF (\ref{Box}) is an example of a biorthogonal ensemble (see e.g.~\cite{Bor99}, \cite[\S 5.8]{Fo10}), and consequently its
 correlation functions are determinantal and so of the form (\ref{D}).
 Moreover, the corresponding correlation kernel $K_N^{r,s}(x,y)$ has the structure
 \begin{equation}\label{Krs1}
 K_N^{r,s}(x,y) = \sum_{l=0}^{N-1} P_l^{r,s}(x) Q_l^{r,s}(y),
  \end{equation}
  where $P_l$ is a monic polynomial of  degree $l$ and $Q_l$ is in the linear span of the functions in the first column and
  $l+1$ rows of the determinant in (\ref{Box}) that furthermore have the biorthogonality property
   \begin{equation}\label{Krs2}
   \int_0^\infty  P_n^{r,s}(x) Q_l^{r,s}(x) \, dx = \delta_{l,n},  \qquad 0 \le l,n \le N - 1.
   \end{equation}
   In the case of (\ref{Box1}) with $s=0$ the explicit form of this biorthogonal system first was derived in \cite{AIK13}.
   A simplified derivation was subsequently given in \cite{KZ13}. In fact the working of  \cite{KZ13} requires only minor
   modification to determine the biorthogonal system for general $r,s$ (for notational convenience we will drop the
   superscripts $r,s$ from $P_n^{r,s}(x), Q_l^{r,s}(y)$ below).
   
 \begin{prop}\label{P4}
 Consider the PDF proportional to (\ref{Box}). The corresponding biorthogonal system as gives rise to (\ref{Krs1})
 and satisfies (\ref{Krs2}) is specified by
  \begin{equation}\label{QBv}
  Q_l(x) = {(-1)^l \over C^{}_l}
   G^{r+1,s}_{s+1,r+1} \Big ({ -(\mu_1 + N), \dots
, - (\mu_s+N), -l  \atop \nu_0,\nu_1,\dots,\nu_r} \Big | x \Big ) 
  \end{equation}
  where $\nu_0:=0$ and
  \begin{equation}\label{CB}
    C^{}_l = (-1)^l \prod_{j=0}^r \Gamma( \nu_j + l + 1) \prod_{p=1}^s \Gamma(\mu_p + N - l),
    \end{equation} 
and
 \begin{eqnarray}\label{QBv1} 
 \lefteqn{P_n(x) = (-1)^n 
  \prod_{j=1}^r  {\Gamma(\nu_j + n +1 ) \over \Gamma(\nu_j + 1)}
  \prod_{l=1}^s {\Gamma( \mu_l+N - n) \over \Gamma(\mu_l+N)}}
  \nonumber \\
 && \qquad  \times
 \, {}_{s+1} F_r \Big ( {-n,1-\mu_1- N, \dots, 1- \mu_s -N \atop
1+ \nu_1, \dots, 1+\nu_r} \Big | (-1)^s x \Big ).
 \end{eqnarray}
 \end{prop}

 \noindent
 {\rm Proof.} \quad
 From the definition of $Q_l(x)$ it belongs to Span$\, \{ I_j(x) \}_{j=1,\dots,l}$ where $I_j(x)$ is given by
 (\ref{MT}). The dependence on $j$ in the latter then allows us to conclude that
 \begin{equation}\label{Qb}
 Q_l(x) = {1 \over 2 \pi i} \int_{c - i \infty}^{c + i \infty} q_l(u) \prod_{j=1}^r \Gamma(\nu_j - u) \prod_{l=1}^s \Gamma(\mu_l+N+u+1) \,
 x^u \, du
 \end{equation}
 for some polynomial $q_l(x)$ of degree $l$. We want to choose $q_l(s)$ so that
 $$
 \int_0^\infty x^k Q_l(x) \, dx = \delta_{l,k} \qquad {\rm for} \quad 0 \le k \le l.
 $$
 
 Proceeding in an analogous way to the workings in \cite{KZ13}, we see by inspection that
 \begin{equation}\label{Qb1}
 {1 \over C^{}_l} \Big ( {d \over dx} \Big )^l \Big ( x^l I_1(x) \Big )
  \end{equation} 
  has the structure (\ref{Qb}). Furthermore, multiplying by $x^k$ and integrating by parts gives zero for
  $0 \le k < l$, while for $k=l$ it gives
  $$
  {(-1)^l l! \over  C^{}_l}  \int_0^\infty x^l I_1(x) \, dx =
   {(-1)^l l! \over  C^{}_l}  \prod_{j=1}^r \Gamma( \nu_j + l + 1) \prod_{p=1}^s \Gamma(\mu_p + N - l),
   $$
   where the equality follows by noting from (\ref{MT}) that $I_1$, after the change of variables $s \mapsto - s$, is an
   inverse Mellin transform. It follows that $Q_l(x)$ is given by (\ref{Qb1}) with $ C^{}_l$ given by (\ref{CB}).
   
   To obtain the form (\ref{QBv}), we note from (\ref{MT}) that (\ref{Qb1}) is equal to the RHS of (\ref{Qb}) with
   \begin{equation}\label{5.7a}
   q_l(u) = {1 \over C^{}_l} (u + l)_l.
   \end{equation}
   Noting that $(u + l)_l = (-1)^l \Gamma(-u)/\Gamma(-l-u)$,
   substituting for $q_l(s)$ on the RHS of (\ref{Qb}) and comparison with (\ref{GGa}) gives (\ref{QBv}).
  
 We now turn our attention to the derivation of (\ref{Qb}). First we can check from (\ref{pFq}) that (\ref{QBv1}) is monic of degree
 $n$. If suffices to check the orthogonality
  \begin{equation}\label{CO}
 \int_0^\infty P_n(x) \tilde{Q}_l(x) \, dx = 0, \qquad l=0,\dots,n-1,
   \end{equation} 
 where $\tilde{Q}_l(x)$ is specified by (\ref{Qb}) with $q_l(u)$ any degree $l$ polynomial. We choose $q_l(s) = s^l$,
 and again proceed in an analogous way to the workings in \cite{KZ13}. The first step is to observe the
 integral representation
 \begin{eqnarray}\label{QBv2} 
 \lefteqn{P_n(x) =  \prod_{j=0}^r  \Gamma(\nu_j + n +1 )
  \prod_{l=1}^s \Gamma( \mu_l+N - n)  } \nonumber \\
  && \times {1  \over 2 \pi i } \oint_{\Sigma}
  {\Gamma(t - n) \over \prod_{j=0}^r  \Gamma(\nu_j + t +1 )
  \prod_{l=1}^s \Gamma( \mu_l+N - t) } x^t \, dt,
  \end{eqnarray}
  where $\Sigma$ is a simple anti-clockwise closed contour encircling $t=0,1,\dots,n$. This follows from 
   the fact that the poles of the integrand occur at $t=0,1,\dots,n$, which are the poles of $\Gamma(t-n)/\Gamma(t+1)$
  (recall $\nu_0 := 0$) and the residue theorem.
  
  To verify (\ref{CO}), we note that $\tilde{Q}_l(x)$, after the change of variables $s \mapsto -s$, is an inverse Mellin
  transformation, and thus
  $$
  \int_0^\infty x^t \tilde{Q}_l(x) \, dx = (t+1)^l   \prod_{j=1}^r  \Gamma(\nu_j + t +1 )
  \prod_{l=1}^s \Gamma( \mu_l+N - t).
  $$
  It follows from this and the integral representation (\ref{QBv2}), upon a change of order of integration, that
  $$
 \int_0^\infty P_n(x) \tilde{Q}_l(x) \, dx  =  { \prod_{j=0}^r  \Gamma(\nu_j + n +1 )
  \prod_{l=1}^s \Gamma( \mu_l+N - n)  \over 2 \pi i}  \oint_{\Sigma} {\Gamma(t - n) (t+1)^l \over \Gamma(t+1)} \, dt.
  $$
 But we know from   \cite{KZ13} that the integral on the RHS vanishes for $ l=0,\dots,n-1$, which is the sought result.
 \hfill $\square$
 
 \medskip
 The form of the PDF (\ref{Box}) is only valid for $s \ge 1$. However the expressions obtained in Proposition \ref{P4} for $Q_l(x)$ and
 $P_n(x)$ are both valid for $s=0$. Indeed, setting $s=0$ gives the expressions obtained in \cite{AIK13} and \cite{KZ13} for the biorthogonal
 system in this case, which were derived for the PDF (\ref{Box1}) with $s=0$. Note in particular that  $P_n(x)$ as specified by 
 (\ref{QBv1}) is precisely the characteristic polynomial (\ref{KP}) with $n=N$, in agreement with general theory \cite{BK03,DF07}. For $s \ge 1$
 we have already remarked that the averaged characteristic polynomial is ill-defined, and in Section \ref{S3} we considered instead the
 averaged characteristic polynomial for the corresponding generalized eigenvalue problem. Comparison of the evaluation of the
 latter (\ref{Prs}) with (\ref{QBv1}) shows very similar structures, although we are unaware of any general theory that predicts the
 precise relationship.
 
 Before we proceed to consider the hard edge limit of the correlation kernel (\ref{Krs1}) as implied by Proposition \ref{P4}, we pause
 to make note of the normalization constant required to fully specify (\ref{Box}) as a PDF.
 
 \begin{cor}
 To be correctly normalized, (\ref{Box}) must be multiplied by
 \begin{equation}\label{K3}
 {1 \over N!} {1 \over \prod_{l=0}^{N-1} C_l},
 \end{equation}
 where $C_l$ is given by (\ref{CB}).
 \end{cor}
 
 \noindent
 {\rm Proof.} \quad
 It follows from the Vandermonde determinant formula that
 $$
 \prod_{1 \le j < k \le N} (x_k - x_j) = \det [ p_{j-1}(x_k) ]_{j,k=1,\dots,N}
 $$
 for $\{ p_l(x) \}_{l=0,\dots,N-1}$ an arbitrary set of monic polynomials, $p_l(x)$ of degree $l$. The polynomial $P_n(x)$ in 
 (\ref{QBv1}) is monic so we can take $p_l(x) = P_l(x)$.
 
  For the determinant in (\ref{Box}) to be unchanged we require the polynomial $q_l(u)$ in (\ref{Qb}) to be monic of degree $l$ in $u$,
  as can be seen from (\ref{MT}). We see from (\ref{5.7a}) that the coefficient of $u^l$ in $q_l(u)$ is in fact required to equal $1/C_l$ for
  (\ref{Krs2}) to hold. Thus for this purpose we must multiply (\ref{Box}) by $1/\prod_{l=0}^{N-1} C_l$. With this done,  (\ref{Krs2})  implies
  the integral over $x_l \in (0,\infty)$ $(l=1,\dots,N)$ is equal to $N!$ (recall (\ref{170})). Thus the factor of $1/N!$ in (\ref{K3}). \hfill $\square$
 
 \subsection{Hard edge limit}
 A remarkable property of kernels for determinantal point processes in random matrix theory is that they admit double contour integrals.
 This is not the result of a general theorem, but rather a feature that has been observed on a case-by-case basis (see e.g.~\cite[Prop. 5.8.3]{Fo10}.
 Moreover, the double contour form of the correlation kernel has shown itself to be well suited to asymptotic analysis (see e.g.~\cite{AvMW13}).
 
A double contour representation of the correlation kernel (\ref{Krs1}) in the case $s=0$ as been obtained in \cite{KZ13}, and this
used to compute the hard edge limit. Using the integral forms of $Q_l(x)$ and $P_n(x)$ given in the proof of Proposition \ref{P4}, the
working of \cite{KZ13} can be readily generalized to the case of general $r,s$.

\begin{prop}
Let  the contour $\Sigma$ be as in (\ref{QBv2}). We have
\begin{multline}\label{bq}
 K_N^{r,s}(x,y)  =  {1 \over (2 \pi i)^2} \int_{-1/2 - i \infty}^{-1/2 + i \infty} du  \oint_{\Sigma} dt \,
   \prod_{j=1}^r { \Gamma(\nu_j + u+1 ) \over   \Gamma(\nu_j + t+1 ) }
    \prod_{l=1}^s { \Gamma(\mu_l+ N - u ) \over   \Gamma(\mu_l + N - t ) } \\
   \times {\Gamma(t - N + 1) \over \Gamma(u - N + 1)} {x^t y^{-(u+1)} \over u - t}.
   \end{multline}
   From this we deduce the hard edge scaled form of the correlation kernel
 \begin{multline}\label{QBv3}
 \lim_{N \to \infty}   {1 \over N^{s+1}} K_N^{r,s} \Big ( {x \over N^{s+1}},   {y \over N^{s+1}} \Big ) =: K_{\rm hard}^{r}(x,y) \\
 =
 \int_0^1   G^{1,0}_{0,r+1} \Big ({ \underline{\hspace{0.5cm}}
 \atop -\nu_0,-\nu_1,\dots,-\nu_r} \Big | u x \Big ) G^{r,0}_{0,r+1} \Big ({ \underline{\hspace{0.5cm}}
 \atop \nu_1,\dots,\nu_r,\nu_0} \Big | u y \Big ) \, du.
 \end{multline}
\end{prop}

 \noindent
 {\rm Proof.} \quad
 For $ Q_l(y)$ we have the integral representation (\ref{Qb}), and for $P_n(x) $ we have the integral representation (\ref{QBv2}).
 After changing variables $y \mapsto - y$ in the former, substitution into (\ref{Krs2}) gives
 \begin{multline*}
 K_N^{r,s}(x,y)  =  {1 \over (2 \pi i)^2} \int_{c - i \infty}^{c + i \infty} du  \oint_{\Sigma} dt \,
   \prod_{j=0}^r { \Gamma(\nu_j + u ) \over   \Gamma(\nu_j + t+1 ) }
    \prod_{l=1}^s { \Gamma(\mu_j + N+1  - u ) \over   \Gamma(\mu_j + N - t ) } \\
   \times  \sum_{k=0}^{N-1} {\Gamma(t - k ) \over \Gamma(u - k)} {x^t y^{-u} \over u - t}.
   \end{multline*}
   
Proceeding as in \cite{KZ13}, we note that  the sum has a telescoping property which allows the evaluation
   $$
     \sum_{k=0}^{N-1} {\Gamma(t - k ) \over \Gamma(u - k)} = {1 \over u - t - 1} \Big ( {\Gamma(t - N + 1) \over \Gamma(u - N)} -
     {\Gamma(t + N) \over \Gamma(u)} \Big ).
 $$
  In the first contour integration, we take $c=1/2$, while in the second we let $\Sigma$ go around $0,1,\dots,n$ with Re$\, t > -1/2$.
  Then $u - t - 1 \ne 0$ along either of the contours. Substituting the telescoping sum    then shows that implied second double
  contour integral vanishes. Changing variables $u \mapsto u + 1$ in the first gives (\ref{bq}).
  
  For the scaled hard edge limit, noting that for large $N$
  $$
  {\Gamma(t - N + 1) \over \Gamma(u - N + 1)} \sim N^{t-u} {\sin \pi u \over \sin \pi t}
  $$
  we see from (\ref{bq}) that
  $$
 \lim_{N \to \infty}   {1 \over N^{s+1}} K_N^{r,s} \Big ( {x \over N^{s+1}},   {y \over N^{s+1}} \Big ) =
 {1 \over (2 \pi i)^2} \int_{-1/2 - i \infty}^{-1/2 + i \infty} du  \oint_{\Sigma} dt \,
   \prod_{j=1}^r { \Gamma(\nu_j + u+1 ) \over   \Gamma(\nu_j + t+1 ) }
{\sin \pi u \over \sin \pi t}      {x^t y^{-(u+1)} \over u - t}
$$
(the implied interchange of the limit and integration can be justified by making use of the dominated convergence
theorem as in \cite{KZ13}).
This is independent of $\{\mu_l\}$ and was obtained in \cite{KZ13} in the case $s=0$.  In \cite{KZ13} this double integral representation was
shown to be equal to the
form on the RHS of (\ref{QBv3}).
\hfill $\square$

\medskip
The hard edge limit of the characteristic polynomial (\ref{Prs}) is particularly simple to compute, it following immediately
from the series definition (\ref{pFq}).

\begin{prop}
Let $\tilde{P}_N^{(r,s)}(\lambda)$ denote the characteristic polynomial (\ref{Prs}) normalized so that the coefficient of the
constant term $\lambda^0$ is unity. We have
\begin{equation}\label{mv}
\lim_{N \to \infty}  \tilde{P}_N^{(r,s)}\Big ( {\lambda \over N^{s+1}} \Big ) = {}_0 F_r  \Big ({ \underline{\hspace{0.5cm}}
 \atop \nu_1+1,\dots,\nu_r+1} \Big | \lambda\Big ) .
 \end{equation}
 \end{prop}

\subsection{Anticipated asymptotic properties and discussion}
Since the case $r=1$, $s=0$ corresponds to classical complex Wishart matrices, the limiting kernel (\ref{QBv3}) must be
equal to the Bessel kernel \cite{Fo93a}. This is verified in \cite{KZ13}. In the case of the Bessel kernel it is well known
that the leading small $y$ form of the global density, which we read off from the Marchenko-Pastur law to be
given by $1/\pi y^{1/2}$, is identical to the leading large $y$ form of the hard edge density as given by the Bessel
kernel \cite[Eq.~(7.33), with $x \mapsto 4x$ and the RHS multiplied by $4$ to account for different constant factor used in the
hard edge scaling]{Fo10}
$$
\rho_{(1)}^{X^\dagger X, {\rm h}}(x) =  \Big ( ( J_a(2\sqrt{x}))^2 - J_{a+1}(2\sqrt{x}) J_{a-1}(2\sqrt{x}) \Big ), \quad a:= \nu_1
$$
(the superscript `h' here and below denotes the hard edge).
As discussed in \cite{GFF05} this corresponds to a matching between asymptotic expansions. Assuming that this
persists for the product complex Wishart matrices, we then obtain from (\ref{B2}) the prediction that for large $x$
\begin{equation}\label{X1}
\rho_{(1)}^{(X_{r,s})^\dagger X_{r,s}, {\rm h}}(x)  \mathop{\sim}\limits^{?}  { \sin \pi/(r+1) \over \pi x^{r/(r+1)}}.
\end{equation}
Numerical evaluation of the ratio of the LHS to the RHS in the cases $r=2$ and $r=3$, and $\{\nu_j\}=0$, gives consistency with this
formula, and shows furthermore that the correction term is oscillatory as the ratio oscillates above and below unity.

Assuming the validity of  (\ref{X1}) it follows that
$$
\lim_{c \to \infty} \int_c^{c + \pi x c^{r/(r+1)}/\sin (\pi /(r+1))}  \rho_{(1)}^{(X_{r,s})^\dagger X_{r,s} \,{\rm h}}(x)  \, dx = x.
$$
Again as with the case $r=1$ \cite[Exercises 7.2 q.2]{Fo10}, this suggests the form of the scaled variables
from hard edge to bulk behaviour,
\begin{multline}\label{X2}
\lim_{c \to \infty}  { \pi  c^{r/(r+1)} \over \sin (\pi /(r+1)) }
 K_{\rm hard}^{r}\Big (c + \pi x c^{r/(r+1)}/\sin (\pi /(r+1)),c + \pi y c^{r/(r+1)}/\sin( \pi /(r+1)) \Big ) \\
  \mathop{=} \limits^{?}  {h(x) \over h(y)}{\sin \pi (x -y) \over \pi (x - y)},
  \end{multline}
where the RHS --- a gauge factor which cancels out of the determinant (\ref{D}) times
the so called sine kernel --- is the bulk scaled kernel for unitary invariant random matrix
ensembles (see e.g.~\cite{Ku11}).  In gathering numerical evidence, to cancel out the gauge
factors we consider the product $K_{\rm hard}^{r}(x,y) K_{\rm hard}^{r}(y,x)$, appropriately scaled.
Here the numerical evidence is less convincing. For
a start with $c$ fixed the accuracy at $x=y=0$ relies on the accuracy of (\ref{X1}) which we already know has oscillatory corrections.
Choosing a value of $c$ for which the scaled product is equal to a value close to unity e.g.~$c=95$ and increasing $y$, we then find 
in the case $r=2$, $\{\nu_j\}=0$ that when tabulated against $(\sin \pi y/ (\pi y))^2$ only  qualitative agreement is
found. For example, the scaled product reaches a value near zero at $y=1.3$ instead of $y=1$.
This  is in distinction to a quantitative agreement found for the same tabulation 
in the case  $r=1$, when (\ref{X2}) is a known to be true.

We are having to resort to numerical computations in relation to (\ref{X1}) and (\ref{X2}) because we don't have command
of the large $x,y$ asymptotics of the integral formula (\ref{QBv3}). We expect that this asymptotic form is given by the replacing
the Meijer G-functions in the integrand by their leading large argument form. 
Generally Meijer G-functions $G_{p,q+1}^{1,p}$ with first parameter in the bottom row zero
are examples of the generalized hypergeometric function ${}_pF_q$
\cite[Eq.~(2.26)]{Fi72}. The first of the Meijer G-functions in (\ref{QBv3}) is of this type, so it can be rewritten
\begin{equation}
 G^{1,0}_{0,r+1} \Big ({ \underline{\hspace{0.5cm}}
 \atop -\nu_0,-\nu_1,\dots,-\nu_r} \Big | u x \Big )  = {1 \over \prod_{j=1}^r \Gamma(1 + \nu_j)} \,
 {}_0 F_r  \Big ({ \underline{\hspace{0.5cm}}
 \atop 1  + \nu_1,\dots, 1 + \nu_r} \Big | -u x \Big ).
 \end{equation}
 The advantage of this expression is that the large negative argument form of  $ {}_0 F_r $ is documented in a clean manner
 \cite[Eq.(16.11.9)]{DLMF13}. Using this formula, and specializing to the case $r=2$, each $\nu_j $ equals zero, we read off
 that for large $x$
 \begin{equation}\label{G1}
 G^{1,0}_{0,3} \Big ({ \underline{\hspace{0.5cm}}
 \atop 0,0,0} \Big | u x \Big )  
 \sim {1 \over \pi \sqrt{3} (ux)^{1/3}} e^{3 (ux)^{1/3} \cos (\pi/3) } \cos \Big (3  (ux)^{1/3}  \sin (\pi/3)- \pi/3 \Big ).
 \end{equation}
 The difficulty is with the second Meijer G-function in the integrand of (\ref{QBv3}).
This is in principle known \cite{Fi72} --- it involves a linear
combination of some explicit elementary functions --- but the practical determination of the scalars is difficult.
Using (\ref{G1}) as a guide, and making use of plots of the function, we conjecture that to leading order for
large $x$
\begin{equation}\label{G1a}
 G^{2,0}_{0,3} \Big ({ \underline{\hspace{0.5cm}}
 \atop 0,0,0} \Big | u x \Big )  
 \mathop{\sim}\limits^{?} {2 \over  \sqrt{3} (ux)^{1/3}} e^{-3  (ux)^{1/3} \cos (\pi/3)} \cos \Big (3  (ux)^{1/3} \sin (\pi/3) - \pi/6 \Big ).
 \end{equation}
Consequently, we expect that for large $x,y$
\begin{equation}\label{G1b}
 K_{\rm hard}^{r}(x,y)  \mathop{\sim}\limits^{?} {2 \over 3 \pi (xy)^{1/3}} \int_0^1  \cos \Big (3  (ux)^{1/3}  \sin (\pi/3)- \pi/3 \Big )
 \cos \Big (3  (uy)^{1/3} \sin (\pi/3) - \pi/6 \Big ) \, {du \over  u^{2/3}}.
 \end{equation}
In the case $x=y$ this does indeed give (\ref{X1}). However comparison of numerical values obtained from this  with the exact
values obtained from (\ref{QBv3}) in the cases $x \ne y$ shows large discrepancies, most likely due both (\ref{G1}) and (\ref{G1a})
breaking down near $u = 0$, which in turn is likely important to the asymptotic form in this case.
This would mean (\ref{G1b})  cannot be used in relation to testing the conjecture (\ref{X2}).

Fortunately, in addition to the integral representation (\ref{QBv3}) for $K_{\rm hard}^{\rm r}(x,y)$ there is also a generalized Christoffel-Darboux form
\cite[Prop.~5.4]{KZ13}. Thus, specialising for convenience to the case that each $\nu_j$ is equal to zero, one has
\begin{equation}\label{CD1}
K_{\rm hard}^{\rm r}(x,y) =
{\mathcal B \Big ( G^{1,0}_{0,r+1} \Big ({ \underline{\hspace{0.5cm}}  \atop 0,\dots,,0} \Big | x \Big )  ,
 G^{r,0}_{0,r+1} \Big ({ \underline{\hspace{0.5cm}}  \atop 0,\dots,,0} \Big | y \Big ) \Big ) \over x - y},
 \end{equation}
 where, with $\Delta_x = x {d \over dx}$,  $\Delta_y = y {d \over dy}$,
 $$
 \mathcal B \, (f(x), g(y)) = (-1)^{r+1} \sum_{j=0}^r (-1)^j (\Delta_x)^j f(x) (\Delta_y)^{r-j} g(y).
 $$
In the special case $r=2$, calling $  G^{1,0}_{0,r+1} $ in (\ref{CD1}) $f(x)$, and calling  $G^{r,0}_{0,r+1}$ by $g(y)$, we see from
(\ref{G1}) and (\ref{G1a}) that
 \begin{align}\label{CD2}
 f(x) & \sim {1 \over \pi \sqrt{3} x^{1/3}} e^{3 x^{1/3} \cos \pi/3} \cos(3 x^{1/3} \sin \pi/3 - \pi/3),  \nonumber\\
 x f'(x)  & \sim {1 \over \pi \sqrt{3}} e^{3 x^{1/3} \cos \pi/3} \cos(3 x^{1/3} \sin \pi/3) \nonumber \\
 (x {d \over dx})^2 f(x) & \sim {x^{1/3} \over \pi \sqrt{3}} e^{3 x^{1/3} \cos \pi/3} \cos(3 x^{1/3} \sin \pi/3 + \pi/3).
 \end{align}
 and 
  \begin{align}\label{CD3}
 g(y) & \sim {2 \over  \sqrt{3} y^{1/3}} e^{-3 y^{1/3} \cos \pi/3} \cos(3 y^{1/3} \sin \pi/3 - \pi/6),  \nonumber\\
 y g'(y) & \sim  -{2 \over  \sqrt{3}} e^{-3 y^{1/3} \cos \pi/3} \cos(3 y^{1/3} \sin \pi/3 -\pi/2) \nonumber \\
 (y {d \over dy})^2 g(y) & \sim {2 y^{1/3} \over  \sqrt{3}} e^{-3 y^{1/3} \cos \pi/3} \cos(3 y^{1/3} \sin \pi/3 -  5\pi/6).
 \end{align}
 Suppose now that $x,y$ are both large, with $x/y \to 1$ as is the case on the LHS of (\ref{X2}). Then we see that 
 \begin{align}\label{CD4} 
 &
 - f(x)  (y {d \over dy})^2 g(y) +  x f'(x)  y g'(y) -  (x {d \over dx})^2 f(x) g(y)  \nonumber \\
 & \qquad \sim {1 \over  \pi} e^{3 x^{1/3} \cos \pi/3} e^{-3 y^{1/3} \cos \pi/3}  \sin \Big ( 3 x^{1/3} \sin \pi/3 - 3 y^{1/3} \sin \pi/3 \Big ).
 \end{align}
 Substituting this in (\ref{CD1}), then substituting the resulting expression in the LHS of (\ref{X2}), we obtain agreement with the RHS
 with $h(x) =  e^{3 x^{1/3} \cos \pi/3}$.

In the application of  classical complex Wishart matrices to quantum transport \cite{SMMP91}, \cite{FH94}, the variance of the conductance is a basic
observable quantity. The conductance is the particular linear statistics of the hard edge eigenvalues $G_\alpha = \sum_{j=1}^\infty 1/(1 + \lambda_j/\alpha)$,
where the scale factor $\alpha$ must be taken as large (formally $\alpha \to \infty$; one requires that the linear statistic be slowly varying on
the length scale of the
spacing between eigenvalues --- see \cite[\S 14.3]{Fo10}).
More generally, with $G_\alpha = \sum_{j=1}^\infty g(\lambda_j/\alpha)$ the task is to compute
\begin{eqnarray}\label{VL}
\lefteqn{\lim_{\alpha \to \infty} {\rm Var} \, G_\alpha  } \nonumber \\
&& := \lim_{\alpha \to \infty} \Big ( \int_0^\infty d \lambda_1   \int_0^\infty d \lambda_2 \,
g(\lambda_1/\alpha) g(\lambda_2/\alpha)  \rho_{(2)}^{T, {\rm h}}(\lambda_1,\lambda_2)+ \int_0^\infty d \lambda \, g(\lambda/\alpha)    \rho_{(1)}^{{\rm h}}(\lambda_1) \Big ) \nonumber \\
&& =  \lim_{\alpha \to \infty} \Big ( - 
 \int_0^\infty d \lambda_1   \int_0^\infty d \lambda_2 \,
g(\lambda_1/\alpha) g(\lambda_2/\alpha)  K_{\rm hard}(\lambda_1,\lambda_2)  K_{\rm hard}(\lambda_2,\lambda_1)  
\nonumber \\
&& \hspace*{8cm}
+ \int_0^\infty d \lambda \, g(\lambda/\alpha)  K_{\rm hard}(\lambda,\lambda) \Big ),  \end{eqnarray}
 where in the first line $ \rho_{(2)}^{T, {\rm h}}(\lambda_1,\lambda_2) :=  \rho_{(2)}^{\rm h}(\lambda_1,\lambda_2) -   \rho_{(1)}^{{\rm h}}(\lambda_1) \rho_{(1)}^{{\rm h}}(\lambda_2)$ denotes the truncated two-point correlation function. A fundamental insight in \cite{Be93a} is that  (\ref{VL}) is determined by the functional
 form of the leading large $x, y$ non-oscillatory form of $ \rho_{(2)}^{T, {\rm h}}(x, y)$,
  \begin{equation}\label{VLa}
 \rho_{(2)}^{T, {\rm h}}(x,y) \sim - {1 \over 4 \pi^2 \sqrt{xy}} {x + y \over (x - y)^2}
  \end{equation}
 (see e.g.~\cite[eq.~(7.75)]{Fo10}) to give that
 \begin{equation}\label{VLaa}
 \lim_{\alpha \to \infty} {\rm Var} \, G_\alpha  = {1 \over 4 \pi^2} \int_{-\infty}^\infty | \hat{g}(k) |^2 k \tanh( \pi k) \, dk, \qquad
 \hat{g}(k) = \int_{-\infty}^\infty e^{ik x} g(e^x) \, dx.
 \end{equation}
 Subsequently (\ref{VLa}) was derived directly from the explicit Bessel kernel formula for $K_{\rm hard}$ \cite{BT93}.
 A problem for future work is to compute $\lim_{\alpha \to \infty} {\rm Var} \, G_\alpha$ in the case of the product Wishart ensemble,
 and to similarly compute the large $x,y$ asymptotic form of $\rho_{(2)}^{T, {\rm h}}(x,y)$.
 
 In the special case $r=2$, $\nu_1 = \nu_2 = 0$, the working (\ref{CD2}) and (\ref{CD3}) substituted into (\ref{CD1}) allows the latter question to be
 answered. Thus we find, upon making use too of the averaging replacements
 $$
 \cos(3 x^{1/3} \sin {\pi \over 3} - \alpha)   \cos(3 x^{1/3} \sin {\pi \over 3} - \beta)  \mapsto \cos(\alpha - \beta)
 $$
 and similarly for the $y$-variable, that the analogue of (\ref{VLa}) in this case is
   \begin{equation}\label{VLb}
 \rho_{(2)}^{T, {\rm h}}(x,y) \sim - {1 \over 6 \pi^2 } {1 + (y/x)^{1/3} + (x/y)^{1/3} \over (x - y)^2}.
  \end{equation}
 
 We conclude with some remarks relating to the occurrence of the kernel $ K_{\rm hard}^{2}(x,y)$ in the work \cite{BGS12}. The latter
 computes the hard edge scaled limit of the correlation kernel for the eigenvalues of the 
 matrix $A$ in the so called Cauchy two-matrix model, which is the joint distribution on positive
 definite matrices $A$ and $B$ proportional to 
\begin{equation}\label{DN} 
 { (\det A)^a (\det B)^b e^{- {\rm Tr} \, A}  e^{- {\rm Tr} \, B} \over \det (A + B)^N}.
 \end{equation} 
 The form obtained in  \cite{BGS12} was shown in \cite{KZ13} to be equivalent to $K_{\rm hard}^{r}(x,y)$ with $r=2$, $\nu_1 = a+b$, $\nu_2 = b$.
 Some understanding of this result can be understood by expressing $A$ as a random matrix product.
 
 First we note that (\ref{DN}) results from integrating over the positive definite matrix $C$ in the joint PDF proportional to
 $$
 (\det A)^a (\det B)^b e^{- {\rm Tr} \, A}  e^{- {\rm Tr} \, B} e^{- {\rm Tr} \, (A + B) C}.
 $$
 Integrating this over $B$ gives that the corresponding joint PDF for $A$ and $C$ is proportional to
 \begin{equation}\label{DN1} 
 { (\det A)^a  e^{- {\rm Tr} \, A ( \mathbb I + C)} \over \det (\mathbb I  + C)^{N+b}}.
 \end{equation}
 Now change variables from $A$ to $D$ according to
  \begin{equation}\label{DN2} 
 D =  (\mathbb I  + C)^{1/2} A (\mathbb I  + C)^{1/2}.
  \end{equation}
 This transforms (\ref{DN1}) to read
 \begin{equation}\label{DN3}
  { (\det D)^a   e^{- {\rm Tr} \, D}  \over \det (\mathbb I  + C)^{2N + a+b}}.
   \end{equation}
  
  We see from (\ref{DN3}) that the matrices $C$ and $D$ are independently distributed.
  Now the PDF $\det(\mathbb I + C)^{2N+a+b}$ is realised by matrices $C = S_1 S_2^{-1}$ with $S_1$ a
  complex Wishart matrix $S_1 = G_1^\dagger G_1$, where $G_1$ is an $N \times N$ standard complex Gaussian
  matrix, and $S_2$ a  complex Wishart matrix $S_2 = G_2^\dagger G_2$, where $G_2$ is an $(N+a+b) \times N$ standard complex Gaussian
  matrix (see \cite{GN99} for the real case). It is furthermore the case that the  PDF 
  $(\det D)^a   e^{- {\rm Tr} \, D}$ is realised by complex Wishart matrices $S_3 = G_3^\dagger G_3$, where
  $G_3$ is an $(N+a) \times N$ standard complex Gaussian. It thus follows from (\ref{DN2}) that $A$ can be constructed
  out of complex Wishart matrices according to
  $$
  A = S_2^{1/2}(S_1 + S_2)^{-1/2} S_3 S_2^{1/2} (S_1 + S_2)^{-1/2},
  $$
  which in turn tells us that the eigenvalues of the matrix $A$ in (\ref{DN}) have the same PDF as the matrix product
  $
  S_3 S_2 (S_1 + S_2)^{-1}.
  $
  Since the Wishart matrices $S_2$ and $S_3$ have hard edge exponents 
  $\nu_1 = a+b$ and $\nu_2 = a$, and analogous to the findings of this paper we don't
 expect the inverse matrix $(S_1 + S_2)^{-1}$ to contribute to the hard edge scaling, it follows that the
 hard edge scaled statistical properties of $A$ can anticipated to be $K_{\rm hard}^r(x,y)$ with $r=2$ and
$\nu_1 = a+b$, $\nu_2 = b$ as is indeed the case.  
 \section*{Acknowledgements}
 This work was supported by the Australian Research Council. Comments on the first draft by G.~Akemann, A.~Kuijlaars and J.~Ispen are
 most appreciated.
 
%\bibliographystyle{amsplain}
%\bibliography{book1}

\providecommand{\bysame}{\leavevmode\hbox to3em{\hrulefill}\thinspace}
\providecommand{\MR}{\relax\ifhmode\unskip\space\fi MR }
\providecommand{\MRhref}[2]{%
  \href{http://www.ams.org/mathscinet-getitem?mr=#1}{#2}
}
\providecommand{\href}[2]{#2}

\end{document}